\documentclass[12pt]{article}
\usepackage{color}
\usepackage{hyperref}
\usepackage{enumerate}
\usepackage{graphicx}
\usepackage[utf8]{inputenc}
\usepackage[english]{babel}
\usepackage{amssymb }
 \usepackage{titling}
\usepackage[usenames,dvipsnames]{xcolor}
\usepackage{placeins}
\usepackage{amsmath}
\usepackage[%
	numbers,sort&compress%
  ]{natbib}
  
\begin{document}

\title{Probabilistic multilayer networks}

\author{%
		Enrique Hern\'andez-Lemus$^{1,2,*}$,
  	Jes\'us Espinal-Enr\'iquez $^{1,2}$\\
  	Guillermo de Anda-J\'auregui $^{1,2,3}$\\
1. Computational Genomics Division,\\ National Institute of Genomic Medicine, M\'exico\\
2. Center for Complexity Sciences, \\Universidad Nacional Aut\'onoma de M\'exico\\
3. C\'atedras Conacyt, \\Consejo Nacional de Ciencia y Tecnolog\'ia, M\'exico\\
Correspondence should be addressed to\\ Enrique Hern\'andez-Lemus: ehernandez@inmegen.gob.mx}

\maketitle 

\begin{abstract}
Here we introduce probabilistic weighted and unweighted multilayer networks as derived from information theoretical correlation measures on large multidimensional datasets. We present the fundamentals of the formal application of probabilistic inference on problems embedded in multilayered environments, providing examples taken from the analysis of biological and financial systems, specifically regulatory cancer genomics and US stock markets. Probabilistic multilayer networks as described here may be useful for mathematical modeling and data analysis of a quite diverse range of multi-context problems; thus possessing a wide range of applicability due to their intrinsic simplicity and generality.\\
Probabilistic multilayer networks; Mutual information; Connectivity patterns in real datasets; Tensor representations; Data analytics
\end{abstract}

\section{Introduction}
In recent times, a wide variety of complex phenomena in the physical, biological and socio-political sciences have become amenable to study, largely due to the progressive abundance of larger, more comprehensive databases. This phenomenon, that has been termed the Big Data revolution, has brought the need to develop more powerful analytical approaches and computational techniques that will enable us to understand at a deeper level the intricate relationships that lie within such large data corpora. Complex networks, in particular, have been extremely successful to provide insight into the structure and function of natural, technological and social systems.\\

However, the full potential of the complex network approach has not been exploited yet. Two main avenues of improvement of the network paradigm have gained interest recently. Recognizing the multidimensional nature of many complex systems, conformed by a multitude of descriptive levels or layers, has led to the development of multilayer network theory. Multilayer networks constitute a solid and powerful approach to the study of complex phenomena \cite{de2013,boccaletti2014}. On the other hand,  the actual hierarchical structure of many complex systems can only be accessed via data generated in high-throughput experiments or empirical observations, that by necessity carry on their own set of biases, noise and other sources of complexity. Hence the development of probabilistic approaches to network inference from large, noisy datasets is also an area of increasing interest.\\

 With this in mind, here we introduce probabilistic weighted and unweighted undirected multilayer networks derived from information theoretical correlation measures on large multidimensional datasets. We will present the fundamentals of the formal application of probabilistic inference problems embedded in multilayered environments, as well as a couple of examples taken from the analysis of biological and financial systems.
 
\section{A probabilistic approach to network inference from massive data}

\subsection{Mutual information networks}

Let $i=\{1,2, \dots, N\}$ and $j=\{1,2, \dots, N\}$, be iid random variables. For each duplex $\mathbb{D}_{ij}= (i,j)$ it is possible to define the mutual information (MI) function $I(i,j)$ as follows \cite{cover2012}:

\begin{equation}\label{MI} I(i,j) = \sum_{i \in {\mathcal I}} \sum_{j \in {\mathcal J}} P(i,j) \, \log{\frac{P(i,j)}{P(i)\,P(j)}}
\end{equation}

Here, ${\mathcal I}$ and ${\mathcal J}$ are the complete sampling spaces associated to the random variables $i$ and $j$ respectively --i.e. the sets of all possible values of $i$, and $j$, within a given (large) experimental data corpus $\Omega$, associated with a general probability triple $(\Omega, {\mathcal F}, P)$. $P(i,j)$ is the joint probability distribution of $i$ and $j$ in $\Omega$, whereas $P(i)$ and $P(j)$ are the marginal probability distributions of $i$ and $j$, respectively. As it is widely known, the mutual information function $I(i,j)$ quantifies the statistical dependence between two given random variables $i$ and $j$ \cite{cover2012}.\\

For each duplex $\mathbb{D}_{ij}$ we can also define the following two functions:

\begin{equation}\label{adj}
A(i,j) = \Theta [I(i,j) - I_0]\cdot \left(1-\delta_{ij}\right)
\end{equation}

and 

\begin{equation} \label{str}
S(i,j) = A(i,j) \, I(i,j)
\end{equation}

$\Theta[\cdot]$ in equation \ref{adj} is Heaviside step function, $I_0$ is a mutual information lower bound or threshold to be determined (for further information on the different methods to set thresholds, see Appendix A), $\delta_{ij}$ is Kronecker's delta. We call $A(i,j)$, the \emph{lower-bounded adjacency function}, for reasons that will become clear soon. Similarly, we call $S(i,j)$ the \emph{lower-bounded weighted adjacency function}.\\

Given the complete set of duplexes $(i,j)$ for all the $N$ random variables, the functions just defined in equations \ref{adj} and \ref{str} can be mapped to two symmetric matrices $\mathbb{A}$ and $\mathbb{S}$ respectively, as follows:\\

Let $A_{ij} = A(i,j)$ and $S_{ij} = S(i,j)$ represent the ij-th elements of the two matrices:
\begin{equation} \label{adjmat}
\mathbb{A} = A_{ij}
\end{equation}
\begin{equation} \label{strmat}
\mathbb{S} = S_{ij}
\end{equation}

$\mathbb{A}$ is, then, the \emph{adjacency matrix} representing the network of relationships between all pairs of random variables (that become the nodes or vertices). An edge exists for each pair of vertices with mutual information higher than a pre-established value $I_0$. The rationale is that variables that are highly statistically-dependent are the ones connected in the network \cite{hernandez2013}. This network represents  the joint probability distribution of the N random variables in the form of a Markov Random Field (MRF). This may result useful later on, since it has been proved that \emph{every MRF has an associated Gibbs measure} \cite{dobruschin1968,hammersley1971,grimmett1973,preston1973}, and this will be relevant for the formal analysis of these systems in the context of statistical physics \cite{binder2018,gandolfi2017,albeverio1975}.\\

 Statistical dependence, as accounted by mutual information measures, has been used as a proxy for \emph{network interactions} for some time now, in particular in the inference of gene regulatory networks from high-throughput experimental data (mostly gene expression) \cite{margolin2006,margolin2006b,tovar2015,espinal2017} as reviewed, for instance, in \cite{hernandez2019}.\\
 
 Gene expression-derived networks have demonstrated being helpful in identification of co-expression patterns between healthy tissue-derived networks and cancer-related networks \cite{espinal2017,dea2019a,dea2019b,dea2019c,garcia2020,dorantes2020}, finding functional modules based on network connectivity patterns \cite{alcala2016,alcala2017,alcala2018,velazquez2019}, or gene clusters grouped by biological attributes such as differential gene expression \cite{zamora2020,garcia2021,andonegui2021}.\\
 
  Similarly, $\mathbb{S}$ is the \emph{weighted adjacency matrix}, also called the \emph{strength matrix}. Here  a weight or strength is given to each edge by the value of the mutual information between its vertices.\\

Equations \ref{adjmat} and \ref{strmat} resume the proposed approach to probabilistic network inference from massive data. It is worth noticing that the case considered here, applies to statistical dependencies between a large number ($N$) of random variables represented on an undirected graph. It does not require any \emph{a priori} knowledge of the structure of interrelationships of those variables. \\

In contrast, if causal relationships are desired, an alternative approach must be used that may be founded on causal reasoning, statistical learning and/or Bayesian approaches \cite{peel2017,peixoto2017,hric2016,casiraghi2017,scholtes2015,peixoto2014}. Nevertheless, applying causal analysis on large multidimensional datasets may require too much additional information in the form of metadata, Bayesian priors, or the use of hidden variables, that quite likely will turn the studies prohibitive due to the large computational burden and complex analytics. More information on alternative methods can be found on the comprehensive review on network reconstruction written by Caldarelli and co-workers \cite{calda2018}.\\

 Although such studies are foreseeable in the future, at this stage we consider that the best way to tackle such problems (causal inference of directed networks) is using the present approach to analyze the large scale system as an undirected, non-causal network inference task. Afterwards we may use the \emph{phenomenology} derived from such study to build intuition and reduce dimensionality, allowing for the use of statistical learning on a reduced problem. Hence, the method just presented may serve for the general analysis of statistical dependencies over multidimensional datasets via undirected (multi) networks, but also as the starting point of exploratory analyses aimed at a causal, directed network inference problem. In the following section we will analyze how to integrate these probabilistic networks when the random variables span over a multidimensional, layered structure.

\subsection{Mutual Information threshold and graph sparsification}

Establishing a numerical value to the parameter $I_0$ in the lower-bounded adjacency function $A(,i,j)$, as defined in equation \ref{adj} is tantamount to set a weight threshold in the associated graph sparsification problem. There are a number of different approaches to the problem of setting a threshold on the weight of the edges so as to discard edges with strength less than a certain value \cite{xiao2018}.\\

In the particular case used here, mutual information $I(i,j)$ is the average log-probability of the statistical independence test condition (or Kullback-Leibler divergence) for two random variables $i$ and $j$. A zero value of $I(i,j)$ implies absolute statistical independence, whereas small values could be related to negligible dependencies that may be due to \emph{sampling errors} and other biases \cite{holmes2018}. Larger values of $I(i,j)$ will imply stronger statistical dependence. Under such circumstances the meaning of the threshold $I_0$ is clear. Slightly dependent pairs of variables will be considered \emph{independent} (thus no edge is drawn into the network between them) whenever $I(i,j) < I_0$.\\

In general, the structure of the resulting network will be strongly dependent on the choice of this threshold, with the exception of community structure due to the high correlation between the degree and the strength of edges on real networks \cite{xiao2018}. Determination of $I_0$ can be made by choosing among a number of ways. For instance, if an accurate measure of the signal-to-noise ratio in the correlations of the data under consideration can be obtained, one possible way to set $I_0$ is by allowing all edges valued above the noise-level. In most practical applications, however, this is not feasible.\\

 Other approaches include the use of statistical sufficiency limits \cite{merchan2016} based on the theory of random constraint satisfaction problems \cite{krzakala2007,gent2001}, the use of bootstrapped edge reshuffling to associate permutation p-values to given mutual information thresholds \cite{margolin2006}, graph sparsification based on analogies with electrical circuits and the \emph{effective resistance} approach \cite{spielman2011}, filtering based on graphs embedded in surfaces of constant genus and the use of topological invariance \cite{tumminello2005} or the use of global null models with invariant strength distributions and global topology \cite{radicchi2011} as well as methods based on multiscale topological properties such as the \emph{backbone} of a complex network \cite{serrano2009}  which preserves edges with statistically significant deviations respect to a pre-assigned local weight null model.
 
\subsection{On statistical sufficiency and conditional independence of pairwise interactions}
Probabilistic multilayer networks inferred via mutual information calculations belong to the class of Markov Random Fields \cite{hernandez2020class} and whenever the joint probability distribution is strictly positive (positive measure everywhere or zero only in a finite number of points, i.e. compact or quasi-compact support) , it may be factorized via Hammersley-Clifford theorem.\\

 Since the measure of a MRF is completely defined through local characteristics, it should satisfy the so-called \emph{pairwise Markov property} (PMP), that is that any non-adjacent variables are conditionally independent given all other variables. PMP may result relevant when defining a particularly useful representation of the joint probability distribution called \emph{Clique factorization}.\\

Let us consider a set of random variables \textbf{X} and a particular configuration $\omega$ in \textbf{X}. $P(\omega)$ is the joint probability for this particular field configuration, then

\begin{equation}
P(\omega) = \prod_{C \in cl(G)} \phi_c(\omega_c)
\end{equation}

\noindent where the sum is over all cliques $C$ in the set of all cliques in the associated graph $cl(G)$. A clique here is a subset of vertices in a graph, such that every two vertices in the clique are adjacent.\\

 Clique factorization (closely related to the associated Gibbs measure of the MRF) is a powerful way to factor the full JPD. It breaks it down to products of probabilities of $0-cliques$  (individual nodes, i.e. marginals), $1-cliques$ (edges, or pairwise joint probabilities),  $2-cliques$, and higher order cliques. As already stated truncating clique factorization at the edge level, implies the assumption of pairwise sufficiency. This is a common practice that applies to a wide variety of systems \cite{merchan2016} and may affect the conditional independence assumption or pairwise Markov property (PMP). There are several ways to try to circumvent this:

\begin{enumerate} 
\item Perform comprehensive data processing inequality (DPI) pruning to preclude the existence of triangles and higher order cliques. It comes with the disadvantage of breaking all loops transforming the MRF into an undirected acyclic graph (a tree).

\item Assuming that the variables are normally distributed, then calculating the inverse covariance matrix, all zero entries are by definition conditionally dependent given all other variables. 

\item Performing regularization either by assuming that the negligible interaction coefficients are small (by applying an $\mathbb(L)_2$ norm, or that the interaction structure matrix is very sparse (via an $\mathbb(L)_1$ norm). The con is that regularization (also known in physics as the \emph{inverse Ising problem} and in machine learning as implementing a \emph{Boltzmann machine}) is a computer intensive task.

\end{enumerate}

\section{Multidimensional probabilistic network integration}
Following the tenets of the multilayer network formalism as developed by De Domenico and co-workers \cite{de2013}, it is possible to describe a layered ensemble of multidimensional interactions, i.e. a \emph{multilayer network}, by means of the so-called \emph{multilayer adjacency tensor}:

\begin{equation} \label{multi}
\mathbb{M} = M_{\beta \tilde{\delta}}^{\alpha \tilde{\gamma}} = \sum_{\tilde{h},\tilde{k} = 1}^{L} \sum_{i,j = 1}^{N} \omega_{ij}(\tilde{h} \tilde{k}) \; \xi_{\beta \tilde{\delta}}^{\alpha \tilde{\gamma}}(ij\tilde{h} \tilde{k})
\end{equation}

The multilayer adjacency tensor $\mathbb{M}$ is the mathematical object that describes properly the intra- and inter- layer connectedness in multidimensional interaction settings. As described in \cite{de2013}, tilded letters refer to any of the $L$ different layers or dimensions, whereas un-tilded letters refer to the $N$ different nodes (that in the present setting represent the set of random variables characterizing the system under study). $\omega_{ij}(\tilde{h} \tilde{k})$ is the associated adjacency  matrix connecting vertex $i$ (which belongs to layer $\tilde{h}$) to vertex $j$ (in layer $\tilde{k}$).\\ 

$\xi_{\beta \tilde{\delta}}^{\alpha \tilde{\gamma}}(ij\tilde{h} \tilde{k})$ is a 4-tensor in the canonical basis for vertices $i$ and $j$ in layers $\tilde{h}$ and  $\tilde{k}$ respectively. The purpose of this 4-tensor is to provide the information on how to properly \emph{embed} the edges within the different layers.\\

The multilayer formalism has been used as a means to integrate several layers of information on different types of underlying network structures \cite{boccaletti2014, kivela2014, de2015struc,de2015rank,de2015ident}. The case of probabilistically-inferred networks built from massive (high-throughput) data is another scenario in which the formal structure of multi-networks provides a solid theoretical foundation for the analysis and integration of complex, multidimensional interactions. Recalling the definitions introduced in the previous section, it is possible to derive expressions for the weighted and unweighted undirected probabilistic multilayer networks as follows.\\

Without losing generality, let the set of random variables $i$ be spanned over a collection of layers (or contexts) $\tilde{h}=\{1,2, \dots, L\}$, and the set of random variables $j$ be spanned over a collection of layers  $\tilde{k}=\{1,2, \dots, L\}$. As stated in \cite{de2013}, the case $\tilde{h} = \tilde{k}$ corresponds to a monolayer setting. An unweighted probabilistic multilayer network, can be obtained by placing $A_{ij} \equiv A_{ij}(\tilde{h} \tilde{k})$ --as given by equation \ref{adjmat}, with vertex $i$ belonging to layer $\tilde{h}$ and vertex $j$ in layer $\tilde{k}$-- as $\omega_{ij}(\tilde{h} \tilde{k})$ in equation  \ref{multi}:

\begin{equation} \label{PUM}
\mathbb{U} = \sum_{\tilde{h},\tilde{k} = 1}^{L} \sum_{i,j = 1}^{N} A_{ij}(\tilde{h} \tilde{k}) \; \xi_{\beta \tilde{\delta}}^{\alpha \tilde{\gamma}}(ij\tilde{h} \tilde{k})
\end{equation}

The probabilistic unweighted multilayer adjacency tensor $\mathbb{U}$, contains all the necessary information to display the unweighted network structure of multidimensional statistical dependencies for a set of $N$ random variables that may exert influence upon each other through $L$ different contexts or layers. \\

While the specific details of the information content of these networks are highly dependent on the choice of the threshold $I_0$, a recent analysis by Yan and coworkers, show that some specific features of the topology of complex networks (most notably, the underlying community structures) are robust to changes in this parameter \cite{xiao2018}. \\

In gene regulatory networks, analogous findings regarding the robustness to changes in threshold values, have been observed in the \textit{physical location} of strongly correlated genes in breast cancer compared to non-cancerous-derived genetic networks \cite{espinal2017}: in breast cancer, genes are more correlated with other genes that belong to the same chromosome and are close between them, the opposite of the case in non-cancer network, where strongest correlations are not dependent on the chromosome location of genes.\\

Much in a similar way, a weighted probabilistic multilayer network is obtained by letting $S_{ij} \equiv S_{ij}(\tilde{h} \tilde{k})$ be used as $\omega_{ij}(\tilde{h} \tilde{k})$ as follows:

\begin{equation} \label{PWM}
\mathbb{W} = \sum_{\tilde{h},\tilde{k} = 1}^{L} \sum_{i,j = 1}^{N} S_{ij}(\tilde{h} \tilde{k}) \; \xi_{\beta \tilde{\delta}}^{\alpha \tilde{\gamma}}(ij\tilde{h} \tilde{k})
\end{equation}

The probabilistic multilayer adjacency tensor $\mathbb{W}$ is just the weighted version of $\mathbb{U}$, representing the same set of multidimensional statistical dependencies, but taking into account the weight or strength of the interactions. The higher the level of statistical dependence between two variables --as given by the mutual information measure--, the stronger the edge between the corresponding vertices.\\

The network structures represented in tensors $\mathbb{U}$ and $\mathbb{W}$  can be used to integrate multidimensional interaction information in order to characterize complex, layered processes. In the following section we will analyze a couple of examples coming from biological and socio-political phenomena, to illustrate the utility of this approach that combines the strength of both, probabilistic data-mining and multilayer network representations.

\section{Probabilistic multilayer graphs}
The multilayer approach has been used to capture both multidimensional contexts in static scenarios and time-dependent network connectivity patterns related to system dynamics \cite{de2013,boccaletti2014,de2015struc}. 

In order to show, how the probabilistic multilayer formalism is able to deal with both types of phenomena, we will illustrate them in this section. We will introduce a multilayered genomic approach to gene regulation through both gene-gene interactions and micro-RNA (miR) post-transcriptional modifications, this will be our example of a multiple context yet static multi-network. 

\subsection{A multi-layer approach to gene regulation in cancer}

Understanding the regulation of gene expression is crucial for the understanding of complex diseases such as cancer. It is known that gene expression may be regulated by the effects of other genes, but also by other mechanisms, such as the influence of miRs, a type of non-coding RNA: sequences in the genome that are not translated into proteins.\\

Based on  gene and miR expression of 86 breast cancer samples available in the Cancer Genome Atlas (TCGA), we used the aforementioned information-theoretical approach to infer gene regulatory networks \cite{drago2017}. Expression values were used to infer Mutual Information using an implementation of the \verb"Infotheo" algorithm \cite{meyer2009package}. With this, we generated a \textit{probabilistic multi-omic network}, containing two layers of distinct classes of biological entities, and three types of probabilistic relationships: among genes, among miRs, and between the two layers.\\

\begin{figure}
\includegraphics[width=\textwidth]{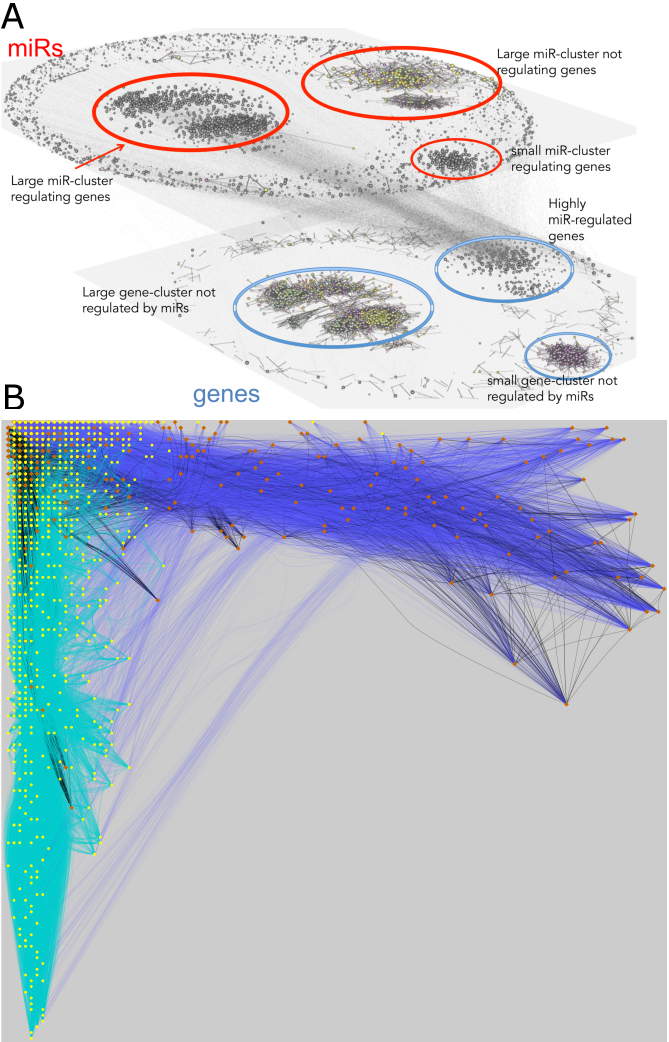}
\caption{\label{fig1} Multilayer network probabilistically inferred from whole genome RNASeq and miRSeq from breast cancer patients in the TCGA collaboration. In A, each layer is represented by 3D squares. Inter-layer edges cross these squares. B, scatterplot-like graph visualization. Here, the nodes are arranged based on their intralayer ( X-axis) and interlayer degree  (Y-axis). The edges are colored by type (gene-gene,  turquoise; miR-miR, blue; interlayer, black).}
\end{figure}

\FloatBarrier

A visualization of this network, rendered using \verb"MuxViz" \cite{de2015} may be found in figure 1A, where each layer contains either genes or miRs. In this figure we can find the existence of isolated gene components (1,2) that show no connection to the miR layer, which implies a regulatory program devoid of miR regulation in the cancer phenotype studied. Meanwhile, there is a large number of genes that show a high number of connections to the miR layer, probably indicating regulation by miRs. Generally, we observe that genes regulated to miRs show no connections to genes that exhibit no miR regulation. Similarly, clusters of interconnected miRs exhibit few links to the gene layer (See Table I).\\

In figure 1B, we provide a scatterplot-like graph visualization, with nodes arranged based on their intralayer and interlayer degree (that is, the number of neighbors in the same layer or in the opposite layer), and edges colored by type (gene-gene, miR-miR, or interlayer). Generally, the highest interlayer degrees are held by miRs, whereas the highest intralayer degrees are held by genes. It can be seen that there is an absence of nodes that simultaneously exhibit high intra and interlayer degree. With this, information transmission between layers will in most cases require more than one step in each layer. This, compounded with the observed isolated components previously described, indicates an incomplete navigability of this network, which may have the biological implication of coexisting parallel regulatory mechanisms involved in the expression of both miR and genes.\\

Gene expression control by the regulatory action of micro-RNAs is a well established phenomenon \cite{baek2008,selbach2008,vasudevan2007}, often involved in cancer \cite{o2010,cho2007}, in particular for breast cancer \cite{drago2017,dea2018}. Most research has been focused on one-to-one --or at most one-to-many-- miR-target gene interactions, with the biological implications of co-regulation of miRs by other miRs being rarely approached; relatively few studies have focused on the concerted action of network-wise regulatory interactions \cite{sales2010,huang2011,drago2017}. The intrinsically multi-scale nature of the regulatory phenomenon makes it ideally suited to be explored from a multinetwork perspective.\\

In this regard, a multilayer network approach to multi-omic regulatory programmes --here, exemplified by a two-omic gene/miRNA regulatory network-- allows both, a clear and concise conceptual characterization of gene regulation not as several disconnected entities, namely gene-gene, gene-miRNA and miRNA-miRNA networks but as a single coordinated object; and also as a tool to characterize the complex phenomenon of gene regulation.\\

Deeper studies on the multilayer nature of this regulatory network may result helpful to develop a mechanistic or semi-mechanistic understanding of the phenomenon, beyond the standard gene-centric view of regulatory interactions. For instance, distinguishing between nodes that are hubs in the gene-gene layer, nodes that are hubs in the miRNA-miRNA layer, and nodes that are hubs in the gene-miRNA interlayer may advance our knowledge of the actual regulatory mechanisms: sequence complementarity based, transcription factor-based, etc.\\

  The same can be said about other network metrics such as multilayer shortest path distributions, edge betweeness, community structure, etc.  Interlayer connections and the underlying network phenomena behind them may provide additional clues about cooperativity among different biological events, though admittedly limited by the pairwise nature of such interactions. Hence, while not picturing the full complexity of transcriptional regulatory phenomena yet, it will advance our understanding of the myriad processes involved under an integrative view.

\FloatBarrier

\begin{table}
\label{Table1}
\begin{small}
\begin{tabular}{lcrcr}
\hline 
\textbf{Network statistic} & & \textbf{gene} & $\;\;\;\;\;\;$ & \textbf{miR} \\ 
\hline 
Number of nodes & $\;\;\;\;\;\;$  & 3750 & & 498 \\ 
\hline 
Maximum degree, intra & & 131 & &  67 \\ 
\hline 
Maximum degree, inter & & 80 & & 427 \\ 
\hline 
Number of edges, intra & & 11189 & & 1629 \\ 
\hline 
Number of edges, inter & & 14064 & & 14064 \\ 
\hline 
\end{tabular}
\end{small}
\caption{Gene-miR Multilayer network statistics}
\end{table} 

\subsection{A multi-layer analysis of stock-market dynamics}
Financial market dynamics pose important restrictions that need to be addressed when designing a hedging approach to investment portfolios. It is generally believed that broad diversification in the most widely diverse funds possible is the most effective ways to hedge a portfolio over the long term. The rationale is that by holding uncorrelated assets and stocks in a portfolio, overall volatility is reduced \cite{larsen1998empirical}. This can be an oversimplified view of the matters that may also prevent investors to move on to more profitable (yet admittedly, more risky) ventures \cite{chiam2013dynamic}. \emph{Uncorrelated} investment portfolios are usually built by assuming that alternative assets may lose less value during a so-called \emph{bear market} (i.e. a market in which securities prices fall 20\% or more from recent highs), so a diversified portfolio will suffer lower average losses \cite{campbell2002increased}. Of course, by the same logic diversified portfolios also offer lower returns during financial turnarounds by not being able to \emph{catch the full tide} \cite{grinold2000active}.\\

An additional concern, often overlooked in the design of secure, diversified portfolios; is the fact that in the complex stock market ecosystem with fast and changing dynamics, second and third-order correlations may not be fully captured by traditional \emph{unidimensional} time series analysis. Multidimensional correlation (and even more statistical dependency) structures of stock market dynamics \cite{machado2011analysis} presents another instance in which probabilistic multilayer networks may offer new insights.\\

To introduce these ideas, we will present a multilayer dynamic network analysis of three of the largest US stock markets: the AMEX, NYSE and NASDAQ. The AMEX (latter known as NYSE American and currently named NYSE MKT) is the oldest of the three \cite{lazonick2007us}. It was formed in the late 18th century, when the American trading market was still developing. Its main goal was to formalize exchange, and to bring rules and regulations to trading practices. The NYSE was formed a little later on, as a competitor specialized in banking operations instead of pure commercial endeavors of the AMEX. Both markets developed to establish as the foundations of the centralized stock market in the United States.\\

 The NASDAQ was formed much later, in the early 1970's as the world's first electronic stock market. Its aim was to regulate the nascent electronic consumer technology markets from a \emph{modern} perspective based on the idea of centralized autonomous control of the stock transactions. In the last three decades these three disparate stock markets (AMEX the commercial, NYSE the banking and NASDAQ the technological) converge onto shared trading options to constitute the largest general financial markets worldwide.

\begin{figure}
\includegraphics[width=\textwidth]{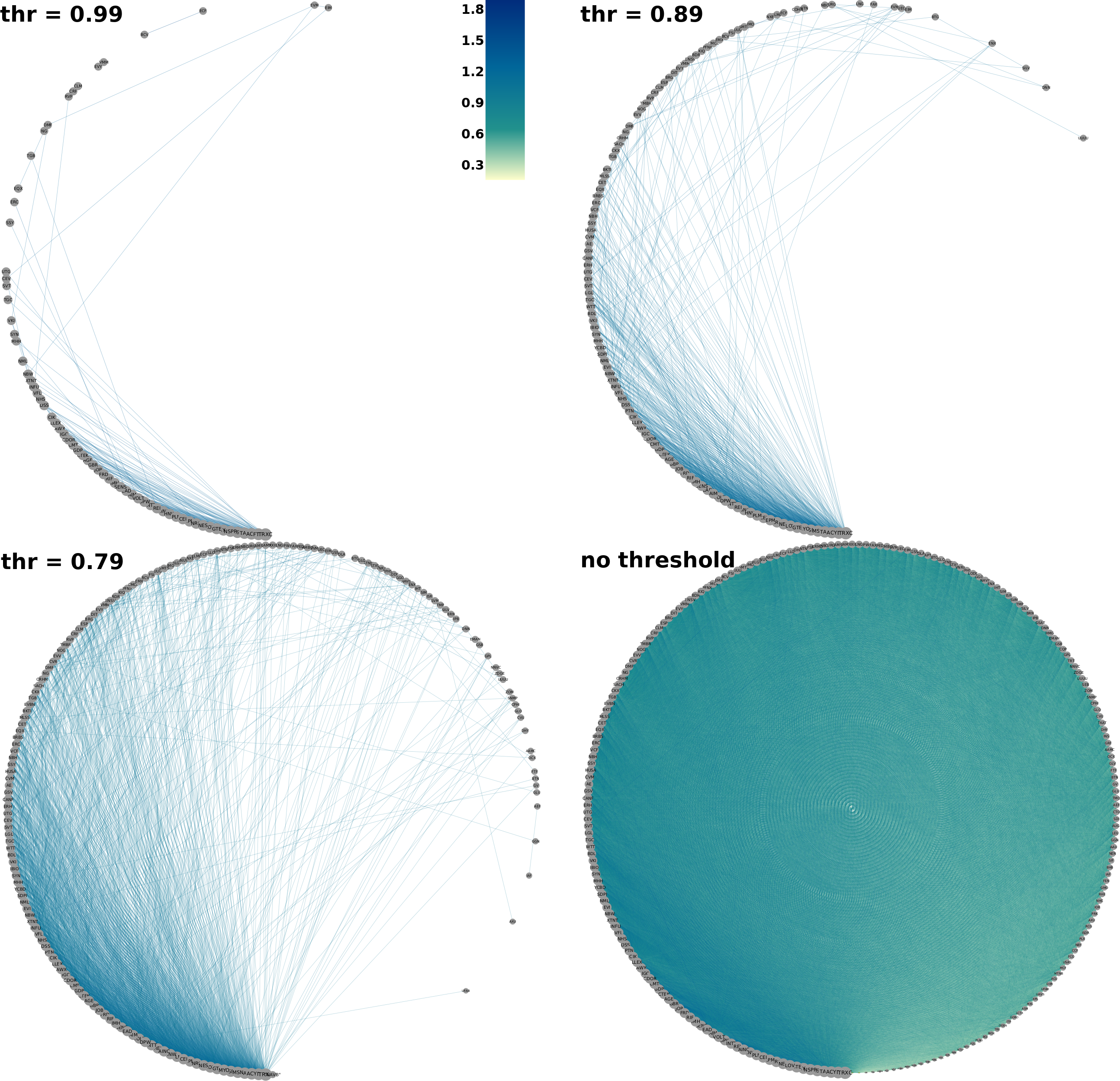}
\caption{\label{fig2} Multilayer networks were probabilistically inferred from US stock exchange companies time series data. Here we show monolayer networks corresponding to the AMEX layer. Networks were pruned according to the set threshold depicted in each figure. Nodes are placed taking into account the average MI value of each node, starting at the bottom of the circle, and following clockwise direction. Links are colored according to the MI values.}
\end{figure}

Every year some 6.5 billion stock market transactions are made in the US financial marketplace, dominated by the AMEX, NYSE and NASDAQ hundreds of thousands of funds and equities \cite{kyle2020microstructure}. The complex dynamics inherent of such an entangled and competitive stock market scenario pose enormous challenges to the financial analysts and portfolio designers worldwide. To present a general panorama of this complex investment ecosystem we have inferred probabilistic multilayer networks from the stock transaction time series of the larger equities from the AMEX, NYSE and NASDAQ markets.\\

In figure 2, the AMEX transaction time series is depicted in the form of a network where nodes represent AMEX equities, and links account for mutual information between two equities time series. In the figure, there are six different samples of the same network with different MI thresholds. Since the AMEX stock market comprises 367 companies, the resulting monolayer network contains $\frac{367^2}{2}=56280$ links. \\

In these representations, nodes are placed in a circle, according to the averaged MI value that each company poses. Hence, the largest averaged MI correspond to those nodes in the bottom center of the  figure, and then decrease following clockwise direction. Maximum MI values in this network are close to 2, meanwhile the minimum values are around 0.03. Links are colored according to the MI value (dark blue to light yellow). \\

 \begin{figure}
\includegraphics[width=\textwidth]{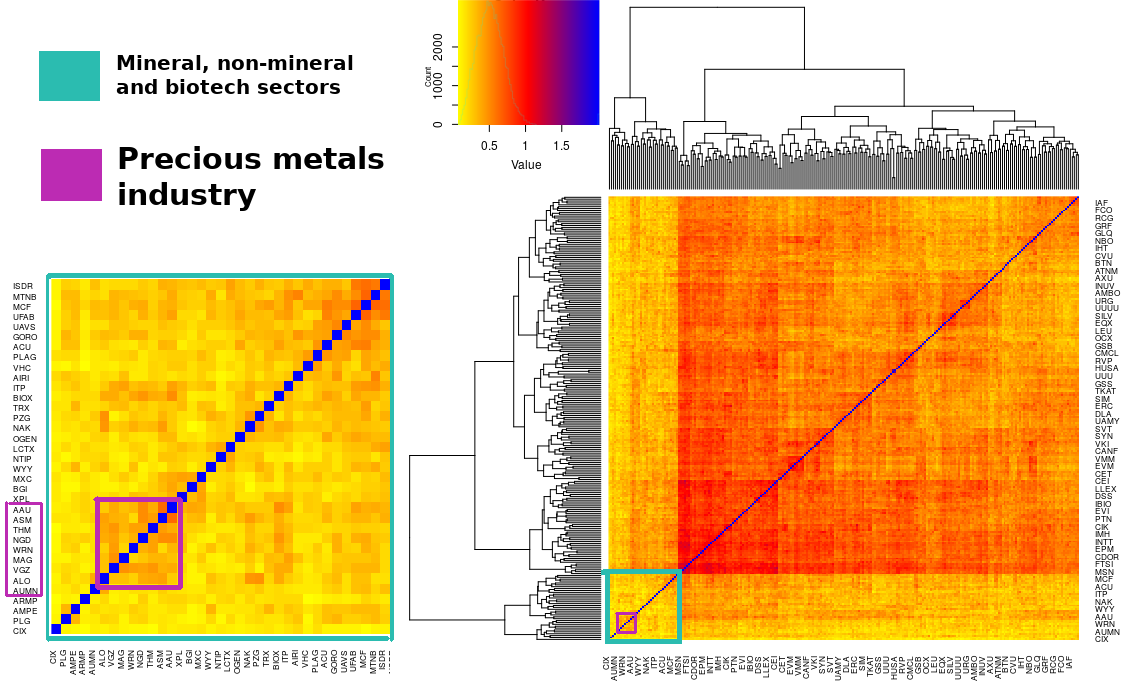}
\caption{\label{fig3} Heatmap of AMEX-AMEX network. In this square matrix, color corresponds to the mutual information between two stocks time series. The left panel is a zoom of the turquoise square of the whole heatmap. The color of squares represents the sector and the industry in which stocks exist.}
\end{figure}

\FloatBarrier

The top left panel of figure 2 shows a pruned network by selecting those interactions with values larger than 0.99. That network contains 66 companies and 202 links. At the top right, a threshold of 0.89 network has 710 interactions with 119 nodes. The bottom right panel shows the whole network with no threshold. \\

 As it can be observed from the placing of the nodes and color of links,  those nodes with the highest MI values between them, are consistently the nodes with largest MI values in general. Additionally, nodes with small MI values are also more grouped between them. \\

 In figure 3, a heatmap constructed by the same AMEX-AMEX network is depicted. As previously mentioned, largest MI values occur between nodes that appear to have always high MI values. The case of low MI values is similar to the previous one, nodes with consistently low MI values tend to cluster together. \\
 
In the right panel of figure 3, the complete interaction network of AMEX stock exchange can be seen. The left panel is a zoom of the turquoise square. In the zoom, companies with low MI values are depicted. The productive sector of those companies is mostly mineral, non-mineral and biotechnology. However, inside the zoom, there is another small cluster of higher MI values (violet square). We may notice that these companies belong to the industry of precious metals.\\

Figure 4 is the heatmap of the NYSE-AMEX interaction network. In this case, since it is  composed of different companies, the heatmap is not a  square matrix. And as in figure 3, clusters  of high and low MI values are formed. It is clear the difference between the bottom left cluster, with red and blue colors, showing higher MI values, and at the same time, top right part of the heatmap contains mostly yellow and orange squares.\\

Figure 4 highlights a richer interaction pattern with the presence of highly correlated regions or blocks. Such blocks are a representation of the multilayered nature of NYSE-AMEX exchange relationships. Closer analysis of such structure may allow to infer inter-market stock operations that may not be obvious of the analysis of the isolated markets.\\

Closer inspection of the multilayered nature of these stock markets may reveal patterns of hidden connections that turn the uncorrelated market hypothesis unable to hold. This is particularly relevant in view that markets are not only correlated via public shared information, but, also quite likely by \emph{biased} information flows coming from both, open multiportfolio holdings, as well as non-disclosed trading speculation. Convoluted trajectories through different layers, in particular, provide a means to account for complex non-linear behaviors whose effects may be observed within a single stock market, but whose origins become untraceable at this level.\\

Multiple stock market trajectories can be further characterized by network statistics at the multinetwork level: multilayer centralities and versatilities, multiple context clustering coefficients accounting for cluster formation in inter-market operations, and so on, as it can be advanced by observing the rich clustering patterns displayed by the heatmap in Figure 4.\\

As in the previous case  --multidimensional gene regulation in cancer--, mining and analyzing the complete multilayer structure, even from this (relatively simple) case of three related stock markets, will constitute work worthy a full line of research. However, we believe that our main goal here, that of outlining (or better, illustrating) the wide scope and broad range of applicability of multilayer networks as probabilistic graphical models in multi-context systems has been accomplished.

  \begin{figure}
\includegraphics[width=9cm]{NYSE_AMEX.png}
\caption{\label{fig4} Heatmap of NYSE-AMEX network. In this square matrix, colors correspond to the mutual information between two stocks time series. The color code is the same than Figure 3.}
\end{figure}

\FloatBarrier

\section{Conclusions}
Probabilistic networks inferred from large multidimensional datasets of empirical evidence, constitute a powerful tool to probe into the intricate multilayer structure of complex systems, being those of physical, biological, social, economic, political or technological origin. Developing a general purpose, theoretical framework to tackle multidimensional probabilistic inference of complex networks is thus an important goal of contemporary (data) science. This challenge can be faced via a multitude of strategies, ranging from Bayesian and causal inference in restricted datasets, to probabilistic machine learning efforts in larger databases.\\

 We consider that a good starting point, of general applicability, can be based on the use of probabilistic inference and information theory to unveil the statistical dependency structure of such complex systems. This is so since such statistical dependence structure  --as given by the joint probability distribution for the (large) set of random variables that best describes the system-- may naturally be cast (via mutual information measures) into a Markov random field that can be represented as a graph in the form of a complex network.\\
 
  A natural extension of the probabilistic graphical model approach just outlined to multidimensional settings is presented here within the formal framework of multilayer network theory. We have used this approach to disentangle and analyze the statistical dependence structure of two quite different complex systems of interest; namely a multidimensional gene regulatory network problem that considers post-transcriptional regulation by micro-RNAs, aside from the common gene regulation program.\\
  
  As is briefly shown, the probabilistic multilayer approach constitutes an effective analytical tool to dig into the complexities of such problems. Due to its general applicability and its relatively low computational complexity, we believe that probabilistic multilayer networks, as presented here, may constitute into a valuable method to analyze multiple context network structure with solid theoretical foundations in Big Data settings.
 
\section*{Acknowledgment}
The present manuscript has been submitted as a preprint with the following accession: \url{https://arxiv.org/abs/1808.07857}

\section*{Author Contributions}
JEE performed calculations, contributed to code, discussed results; GDJ performed calculations, contributed to code, discussed results; EHL developed the theoretical approach, performed calculations, wrote draft manuscript. All authors revised and approbed the manuscript.

\section*{Funding}
This work was supported by the Consejo Nacional de Ciencia y Tecnología [SEP-CONACYT-2016-285544 and FRONTERAS-2017-2115], and the National Institute of Genomic Medicine, M\'exico. Additional support has been granted by the Laboratorio Nacional de Ciencias de la Complejidad, from the Universidad Nacional Autónoma de M\'exico. EHL is recipient of the 2016 Marcos Moshinsky Fellowship in the Physical Sciences. 

\bibliography{apssamp}

\end{document}